\documentclass[fleqn,12pt,twoside]{article}
\usepackage{espcrc1}

\usepackage{graphicx}

\newcommand{\AmS}{{\protect\the\textfont2
  A\kern-.1667em\lower.5ex\hbox{M}\kern-.125emS}}

\title{Coupling nuclear reaction rates with temperature in explosive   
       conditions}

\author{Domingo Garc\'\i a-Senz
 \address{Dpt. de F\'\i sica i Enginyeria Nuclear. UPC. Jordi Girona 1-3,\\  
      M\`odul B5, 08034 Barcelona, Spain
            }\address{Institut d'Estudis Espacials de Catalunya, Gran Capit\`a 
	       2-4, 08034 Barcelona}
            and Rub\'en M. Cabez\'on G\'omez$^{\rm {\ a}}$}

\begin{document}

\maketitle

\begin{abstract}
We present a straightforward integration method to compute the abundance and temperature evolution in explosive scenarios. In this approach the thermal equation is implicitely coupled with chemical equations in order to avoid instabilities and ensure a gentle transition from the normal combustion regime to the quasi (QSE) and complete nuclear statistical equilibrium (NSE). Two nuclear networks, with 14 nuclei ($\alpha$-network) and 86 nuclei (including protons and neutrons) respectively, have been considered. The scheme is suitable to cope with a variety of explosive burning regimes.

\end{abstract}

\section{Introduction}

Violent phenomenae in astrophysics are often related with the sudden release of nuclear binding energy. In some cases, when material is very degenerate, such release is practically isochoric and we refer it as explosive ignition. In other situations there is some expansion and the evolution takes place at nearly constant pressure and high temperature. In both combustion regimes, however, the nuclear system evolves from the initial composition to a final state characterized by the coexistence of hundreds or even thousands elements in chemical equilibrium. From the point of view of stellar modelization, different burning regimes translates into a variety of subroutines in order to properly handle all combustion stages. It is frequent to use a second order implicit scheme to integrate the chemical equations and assume NSE wherever density and temperature are high enough. While this approach has proven useful in many situations, it would be worthwile an integration algorithm suitable to handle rapid combustion, including the limiting cases of QSE and NSE, fast and stable enough to keep large time-steps. A possible solution is to couple molar fractions with temperature through the energy or entropy equation as proposed by E.M\"uller \cite{m86}. Taking a small network of 13 nuclei, from $\alpha$ to $^{56}Ni$, he showed how the inclusion of temperature effects was decisive to avoid the numerical oscillations which otherwise show up during and after the relaxation to the NSE regime. In this paper we extend the previous work of M\"uller considering a larger nuclear network and providing some more test cases. In addition we also propose a slightly different integration method which is very robust and straightforward to implement. Nowadays, the effort to write the proposed numerical scheme for large networks (even 'kilo-nuclide' networks) can take great advantage of libraries of theoretical reaction rates recently published \cite{rt00}.

\section{Description of the method}

First of all the chemical equations are written in their implicit form and linearized taking $Y_i^{n+1}=Y_i^n+\Delta Y_i$ as described in \cite{at69}. In addition we also take $T^{n+1}=T^n+\Delta T$ and make a first order Taylor expansion of nuclear rates with respect the temperature. For instance, for photodesintegration rates we take $\lambda_i^{n+1}=\lambda_i(T^n )+\partial\lambda_i/\partial T\vert_{T^n}\Delta T$. Neglecting second order terms like $\Delta Y_i\Delta Y_k$ or $\Delta Y_i\Delta T$ the discretized system of equations reads:

$$\displaystyle{{\Delta Y_i\over{\Delta T}}=\sum_{k, l}r_{kl} Y_l^n\Delta Y_k-\left({\sum_jr_{ij} Y_j^n}\right)\Delta Y_i+\sum_m\lambda_m\Delta Y_m-\lambda_i\Delta Y_i+\sum_{k,l}r_{kl} Y_k^n\Delta Y_l}$$$$\hskip 7mm\displaystyle{-\sum_{i,j}r_{ij} Y_i^n\Delta Y_j+\left({\sum_{k,l}r_{kl}^{\rq} Y_k^n Y_l^n-\sum_jr_{ij}^{\rq} Y_i^n Y_j^n+\sum_m\lambda_m^{\rq} Y_m^n-\lambda_i^{\rq} Y_i^n}\right)\Delta T}$$$$\hskip -34mm\displaystyle{+\sum_{k,l}r_{kl} Y_k^n Y_l^n-\sum_jr_{ij} Y_i^n Y_j^n+\sum_m\lambda_m Y_m^n-\lambda_i Y_i^n}\eqno(1)$$

\noindent
where $r_{ij}=\rho Na\langle\sigma,v\rangle_{ij}$ stands for particle reactions, $\lambda_i$ is for photodesintegrations, an upper comma means to take the temperature derivative and the remaining symbols have the usual meaning. In addition we need an extra equation to close the system. We took the energy equation, which for an adiabatic process can be written as follows:\medskip

$$\displaystyle{\sum_iBE_i\Delta Y_i-\left\{{\left({\partial U\over{\partial T}}\right)^n-{\Delta\rho\over{\rho^2}}\left({\partial P\over{\partial T}}\right)^n}\right\}\Delta T=-T^n{\Delta\rho\over{\rho^2}}\left({\partial P\over{\partial T}}\right)^n}\eqno(2)$$\medskip

Eventual diffusion terms should be added to the independent terms to the right of Equation 2. This method can be applied to arbitrarly large networks. The computational effort to integrate the system with and without thermal coupling is practically the same.

\subsection{Tests realized}

The nuclear evolution of dense plasmas under extreme thermal conditions involve a large quantity of nuclei. However, in current multidimensional simulations of supernovae and others explosive events the size of the network is strongly constrained by computational efficiency. Small networks of around ten nuclei are used in this case, which approximately give the correct nuclear energy rate. Thus, we have checked the method with two networks of different size: an $\alpha$-chain of 14 nuclei from $\alpha$ to $^{60}Zn$ (including $^{12}C$, and $^{16}O$ binary reactions) and a network of moderate size with 86 nuclei from protons and neutrons to $^{60}Zn$. Nuclear rates were taken from \cite{rt00} and the sparse matrix solver was that of \cite{paa87}.\par
The following cases of astrophysical interest were successfully calculated: a) isochoric combustion of an initial fuel composed by $^{12}C + ^{16}O$ at $\rho = 10^9 g.cm^{-3}$ followed by an adiabatic expansion in the hydrodynamical time (Figure 1, upper right), b) photodesintegration of a system composed of $^{56}Fe$ at constant density of $10^9 g.cm^{-3}$ (bottom left) and, c) isobaric propagation of a nuclear flame in a medium composed by $^{12}C + ^{16}O$ and $\rho = 10^9 g.cm^{-3}$ (bottom right). As a comparation, the isochoric burning case was also calculated without thermal coupling. In this case the oscillations of both, abundances and temperature, are clearly visible when NSE is approached (upper left, using 14 nuclei).\par

\begin{figure}[!t]
\includegraphics[width=18pc]{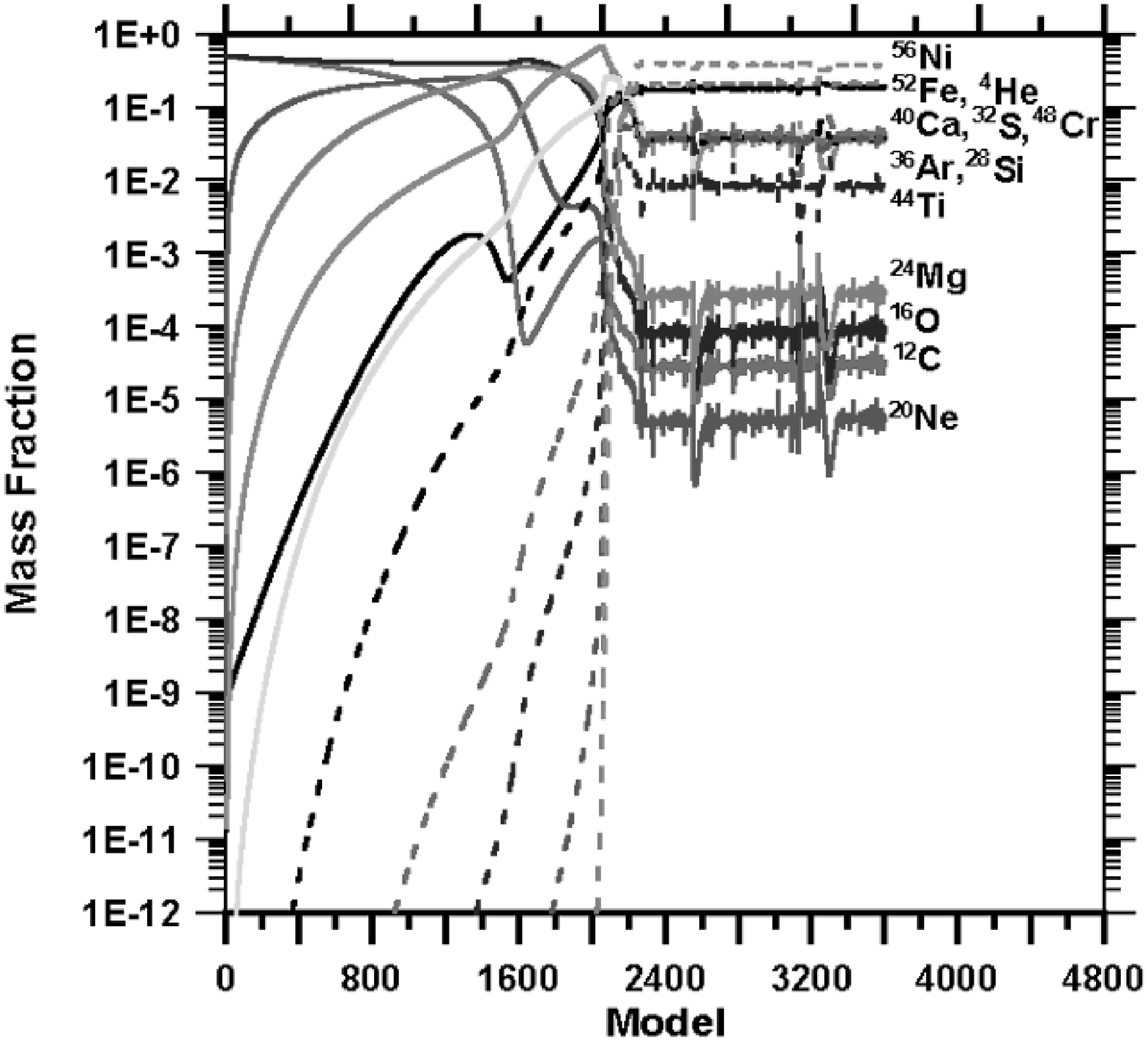}
\includegraphics[width=18pc]{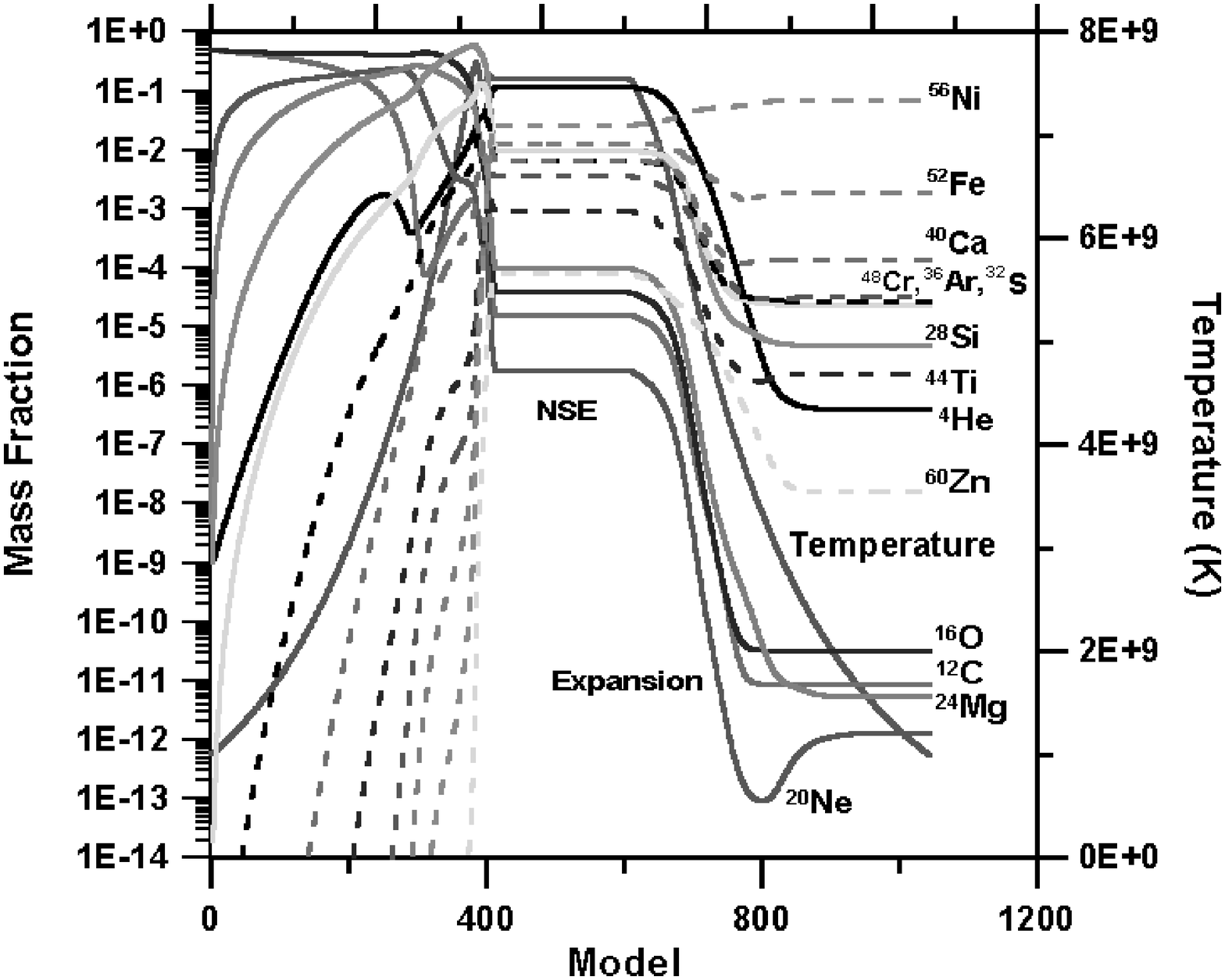}
\par
\hspace{\fill}
\includegraphics[width=18pc]{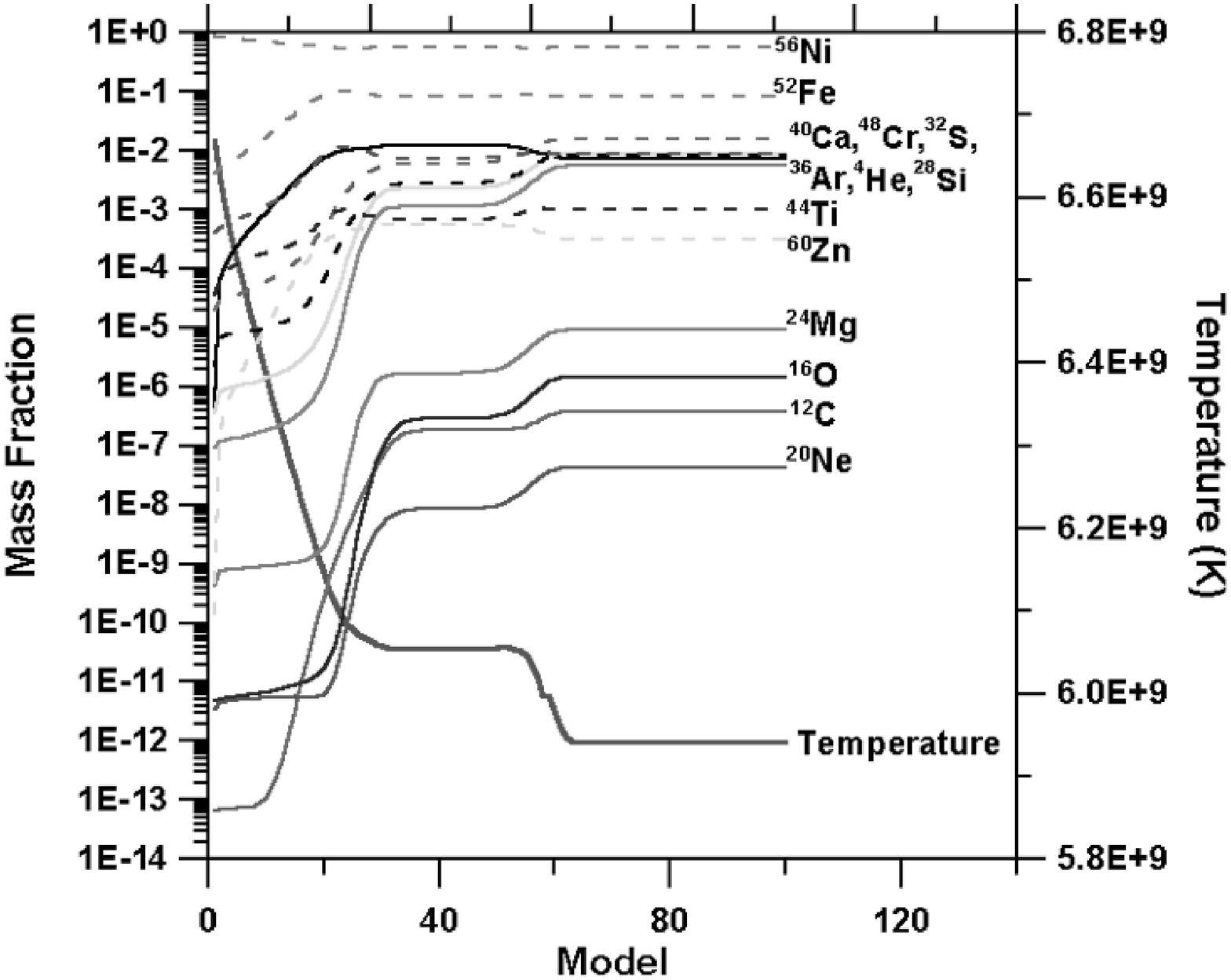}
\includegraphics[width=18pc]{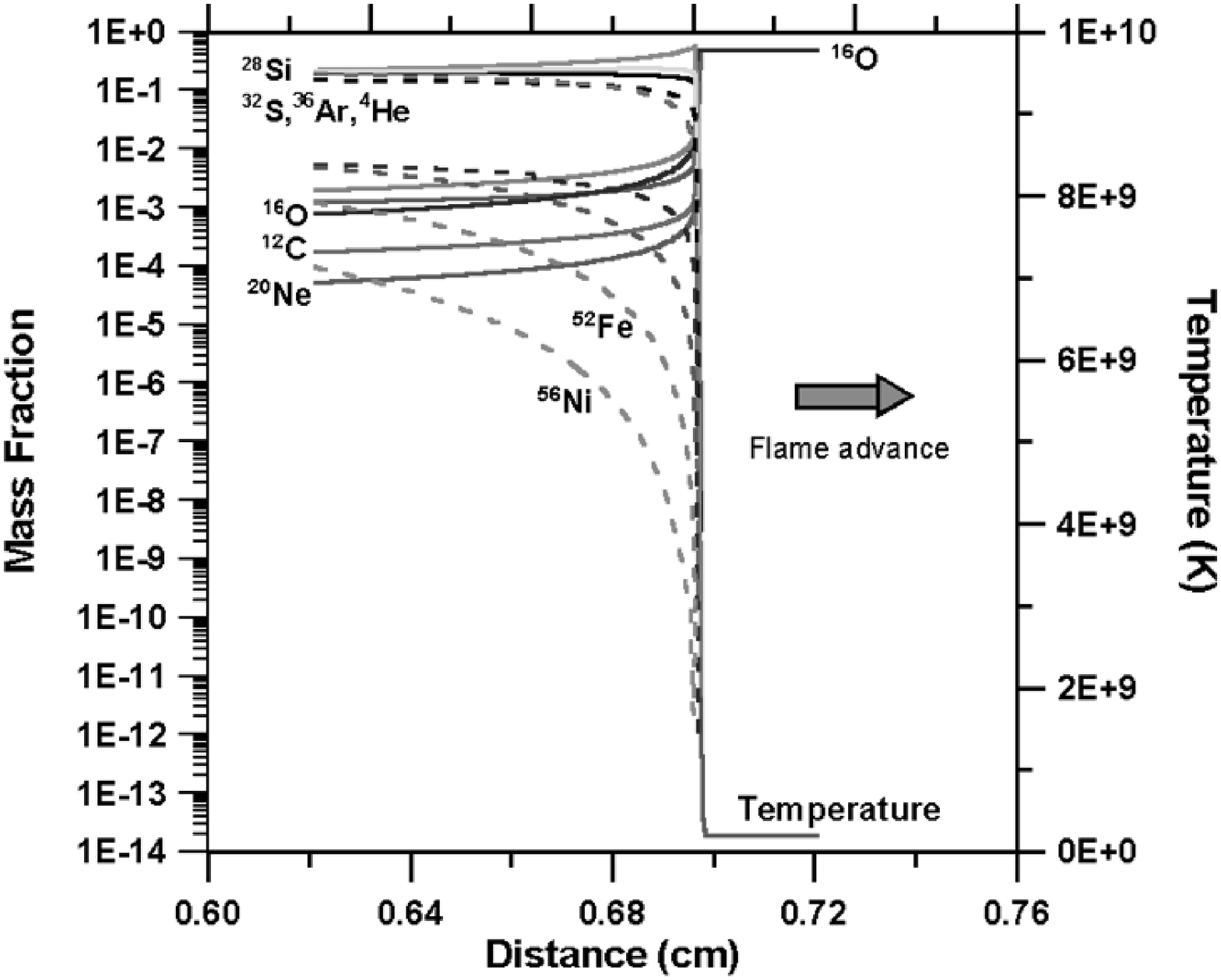}
\begin{minipage}[t]{130mm}
\caption{Nuclear combustion for the four cases discussed in the text.} 
\end{minipage}
\end{figure}

Fortunately, once the thermal coupling was included the QSE and the NSE regimes were smoothly achieved with no fluctuation in all cases, allowing the scheme to take longer time-steps in these  phases. More details about the performance of the method will be presented elsewhere \cite{cg03}.\par
This work has benefited from the MCYT grants EPS98-1348 and AYA2000-1785 and from the DGES grants PB98-1183-C03-02.

\end{document}